# Design of Paper Robot Building Kits

**Ruhan Yang**

ATLAS Institute, University of Colorado Boulder

ruhan.yang@colorado.edu

**Ellen Yi-Luen Do**

ATLAS Institute, University of Colorado Boulder

ellen.do@colorado.edu

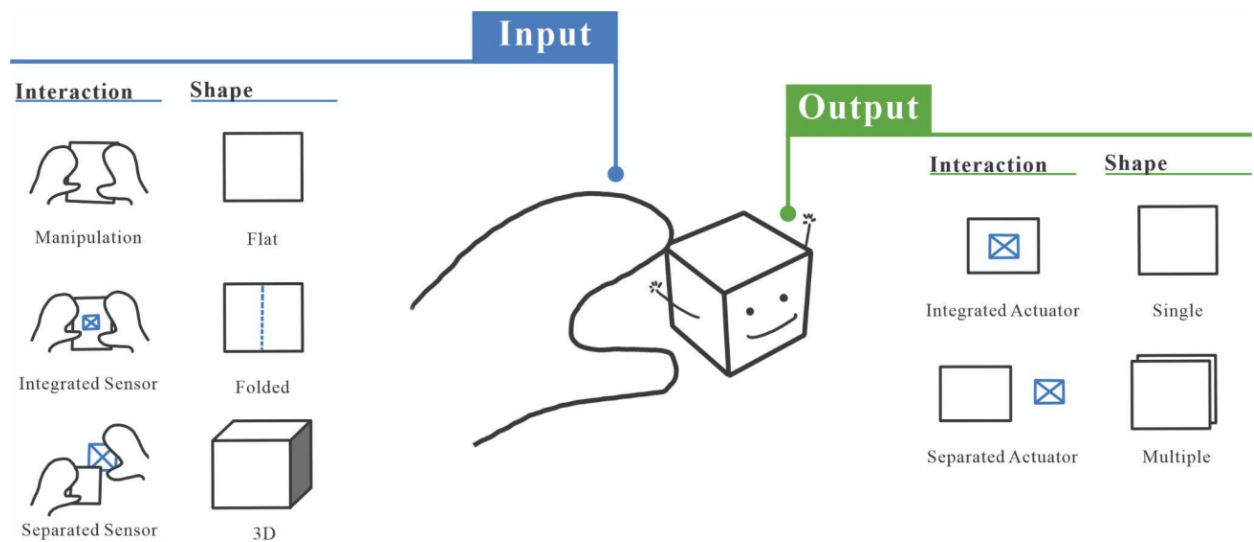

Figure 1: Design space of paper robots

Building robots is an engaging activity that provides opportunities for hands-on learning. However, traditional robot-building kits are usually costly with limited functionality due to material and technology constraints. To improve the accessibility and flexibility of such kits, we take paper as the building material and extensively explore the versatility of paper-based interactions. Based on an analysis of current robot-building kits and paper-based interaction research, we propose a design space for devising paper robots. We also analyzed our building kit designs using this design space, where these kits demonstrate the potential of paper as a cost-effective material for robot building. As a starting point, our design space and building kit examples provide a guideline that inspires and informs future research and development of novel paper robot-building kits.

# 1 INTRODUCTION

With the popularity of STEM education, educational robots have become a part of our lives. Today, there are robot-building kits for people of all ages and skill levels. Conventional robot-building kits are usually made with plastic as it could be shaped in various forms and colors [Freinkel, 2011]. However, those unsustainable plastic components need a specific fabrication process to mold, which makes it costly and challenging to be customized by end users. To improve accessibility, sustainability, and foster creativity, we propose designing paper robot-building kits.

Building robots out of paper offers a unique solution to this issue. Since the 1990s, researchers have explored paper-based interactions. These efforts range from the integration of paper with virtual reality [e.g., Wellner, 1991] to the movement of the paper itself [e.g., Wrensch, 1998]. These studies have provided a solid foundation for the application of paper-based tangible interactions in everyday life. By applying those techniques, we can achieve more creative and expressive designs in robot building without the constraints of cost and fabrication capabilities. The development of paper robot-building kits could also increase public interest in paper-based interaction and promote research and progress in the field. This approach to robot building not only has the potential to change the way we build and design, but it also makes the world of human-computer interaction more accessible and open to all, thus opening up new possibilities for education, entertainment, and more.

However, despite its potential, the design space for paper robots remains largely uncharted. To better apply paper-based interaction to robot building, we surveyed 30 robot-building kits and 30 paper-based interaction studies, identified key design elements, and synthesized them into the following design space of paper robots (see Figure 1). First, we divided all interactions into input and output categories. Input interactions can be achieved through manipulation, integrated sensors, or separated sensors, and appear as flat, folded, or 3D shapes. Output interactions can be implemented by integrated or separated actuators and appear as single or multiple pieces. This design space integrates insights from robot building kits and paper-based interaction research, laying the groundwork for further exploration and innovation in this emerging field.

In this paper, we also showcase several paper robot-building kits to highlight the potential and versatility of paper robot techniques. These example kits provide insight into the design and fabrication possibilities of paper robots, and will serve as inspiration for those who wish to explore this field. We presented these designs to the public, and the positive feedback received emphasizes the feasibility of paper robots and the interest in the future development of this field. The results suggest the potential of paper robot-building kits to play a popular and influential role in the fields of robotics, engineering, crafts, and the arts. By providing the design space and example designs, we aim to motivate others to explore this exciting area and use the information provided in this paper to guide and direct their own work.

In the following, we first introduce the concepts related to paper robot-building kits. We then review the state of the art in robot-building kits and paper-based interaction research, define their features, then describe the design space of the paper robots. This is followed by a demonstration of several building kit designs and an analysis of them based on the design space. Finally, we discuss future research opportunities.

# 2 SCOPE AND METHOD

## 2.1 Scope and Definitions

2.1.1 *Robots and Robot-building Kits*. The scope of robots discussed in this paper is smaller-sized personal robots that can perform some simple movements, including interaction with people or the environment. The building kit discussed in this paper refers to a collection of components that can be assembled to form a certain structure.

When searching using "robot building" and "robot building kit" as keywords, we found that most robot-building kits are intended for education and entertainment. These robots are smaller and have more limited functionality, so they do not include industrial-grade materials or structures; users focus more on building and designing, and rarely require the robot to complete a specified task or complex movement. Rather, these robots play a role in developing hands-on skills, STEM education, and entertainment. Therefore, all the robot-related work we discuss in this paper has an educational or entertainment focus, and we do not cover any specialized robot-building kits or systems (such as Roboteq and ArduPilot). Moreover, we consider STEM building platforms that are provided with parts only such as Arduino and LittleBits as systems of parts, so they are not considered as robot-building kits in this paper.

2.1.2 *Paper and Paper-based Interactions*. Our discussion covers all types of paper and its derivatives, including paper of different weights, paper with different coatings, cardboard, etc. There has been some work discussing paper robots. Some use paper structures and retain the circuitry and electronics of traditional robots, as in [Dibia, 2017]; others have used the properties of paper to complete a functional structure that replaces traditional electronic components, such as [Ryu, 2020]. All these works present us with cases of paper interaction design use in robot building. In this paper, we consider paper-based interaction as a broader scope that includes any interaction that involves the use of paper, regardless of the technology used. We classify these interactions into three types: augmented reality-based interactions (e.g., using projections [Wellner, 1991], sound [Back, 2001], or AR markers [Zheng, 2020]); paper circuits [e.g., Coelho, 2009]; and paper movement [e.g., Saul, 2010].

## 2.2 METHOD

The design space of the paper robot, as well as the designs of the robot-building kits, are derived from a review of the current robot-building kits and paper-based interactions research. In order to collect a representative set of related work, we investigated the commercial market and the research and maker communities.

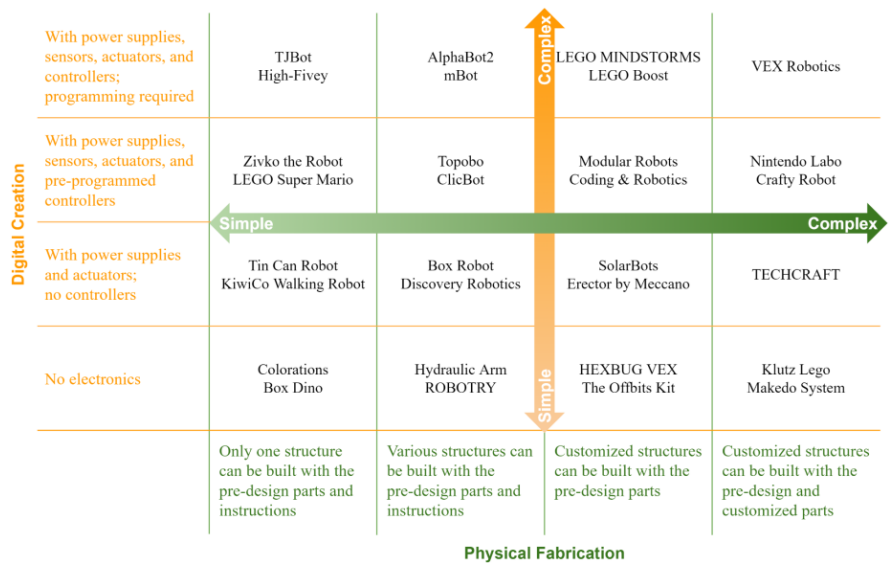

Figure 2: Complexity distribution of robot-building kits

For the robot-building kits, we first surveyed stores including Target and Walmart, as well as online e-commerce sites Amazon and eBay. We searched for "robot building", "robot kit", "robot building kit", "robot building toy", "STEM building kit", "paper robot", "craft robot", and "DIY robot". We then searched Instructables, Pinterest, Maker Faire, YouTube, TikTok, and Etsy, using the same keywords, to learn what is going on in the maker community. Finally we investigated the research community by searching the ACM Digital Library, IEEE Xplore, Springer, and ResearchGate. Based on these results, we also investigated the source of these designs by searching for the company, publisher, and author of them. We identified over 1,000 robot-building kit designs in the first round. From them, we excluded designs that were out of scope such as the more professionally-focused systems and those that did not meet the interaction requirements, such as model robots intended only for decoration. Finally, we summarized the remaining designs, organized them by the way they were physically built and digitally created, and the level of difficulty involved in building them. We finally selected 30 representative designs for detailed analysis. Figure 2 shows the distribution of these building kits in terms of complexity of the digital creation and physical build.

For paper-based interactions, we searched for "paper craft", "paper interaction", "paper mechanisms", "paper computing", "paper circuits", and "origami". We collected 176 papers and grouped them by their interaction designs. For papers that addressed different versions of the same project, we kept only the most recent one; and for different projects that used a similar interaction design, we kept only the earliest one. In the end, we selected 30 representative papers/projects for further review.

Although we have provided an in-depth review into both robot-building kit designs and paper-based interaction research, our list of relevant works is not an exhaustive list of these two fields. Our focus is on the expansion of the design space, so we are interested in the uniqueness of these efforts in terms of how interactions are performed. We first analyzed the designs of the robot building kits and collected the interaction elements included in each kit. We then reviewed paper-based interaction studies and summarized the interaction modalities they proposed. Next, we analyzed and summarized all the interactions and identified the dimensions of the design space. Based on this design space, we analyzed the five building kits that we designed. These building kits were developed to test the feasibility of the concept of building robots out of paper. We deployed these kits in public events, where we observed people working on them.

## 3 INTERACTION DESIGNS FROM ROBOT-BUILDING KITS

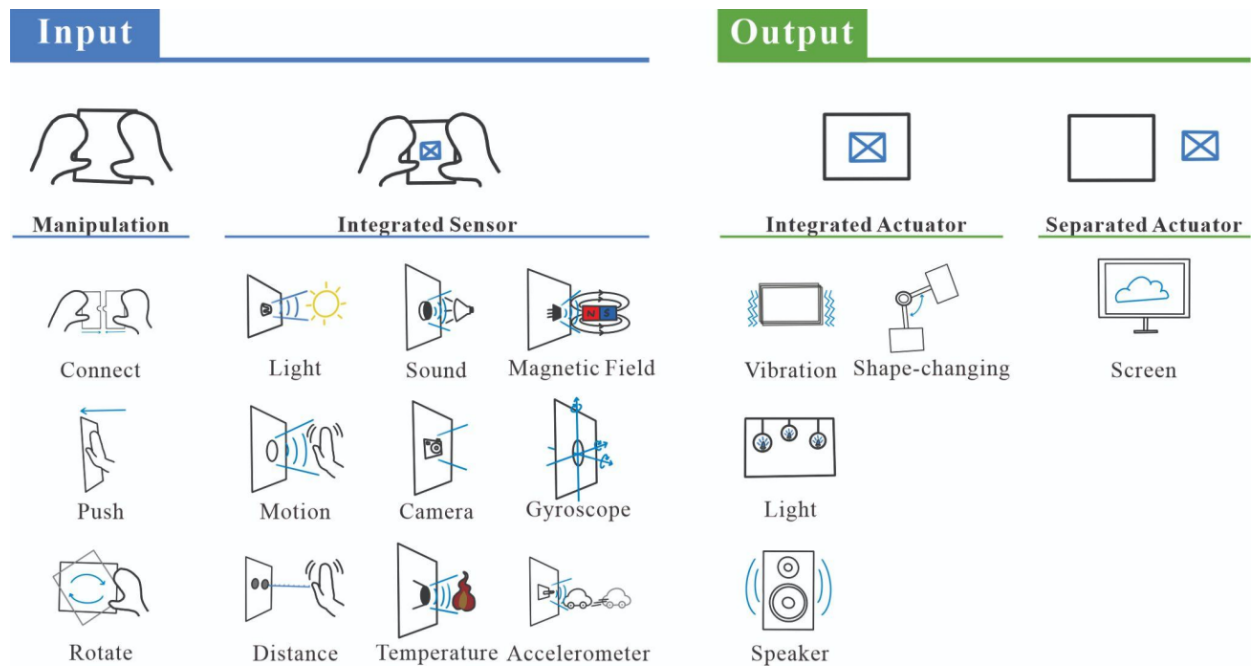

Figure 3: Input and output modalities of robot-building kits

In this section, we delve into the interaction design elements present in robot-building kits. We categorize the input and output modalities of these kits, and we identify two input types: 1) manipulation, and 2) integrated sensors. For the outputs, we propose two output types: 1) integrated actuators, and 2) separated actuators. With these dimensions, we can map the interactions of robot-building kits into an initial design space (Figure 3). In Figure 4, we provide a list of these interactions, indicating the specific kit from which each interaction originates.

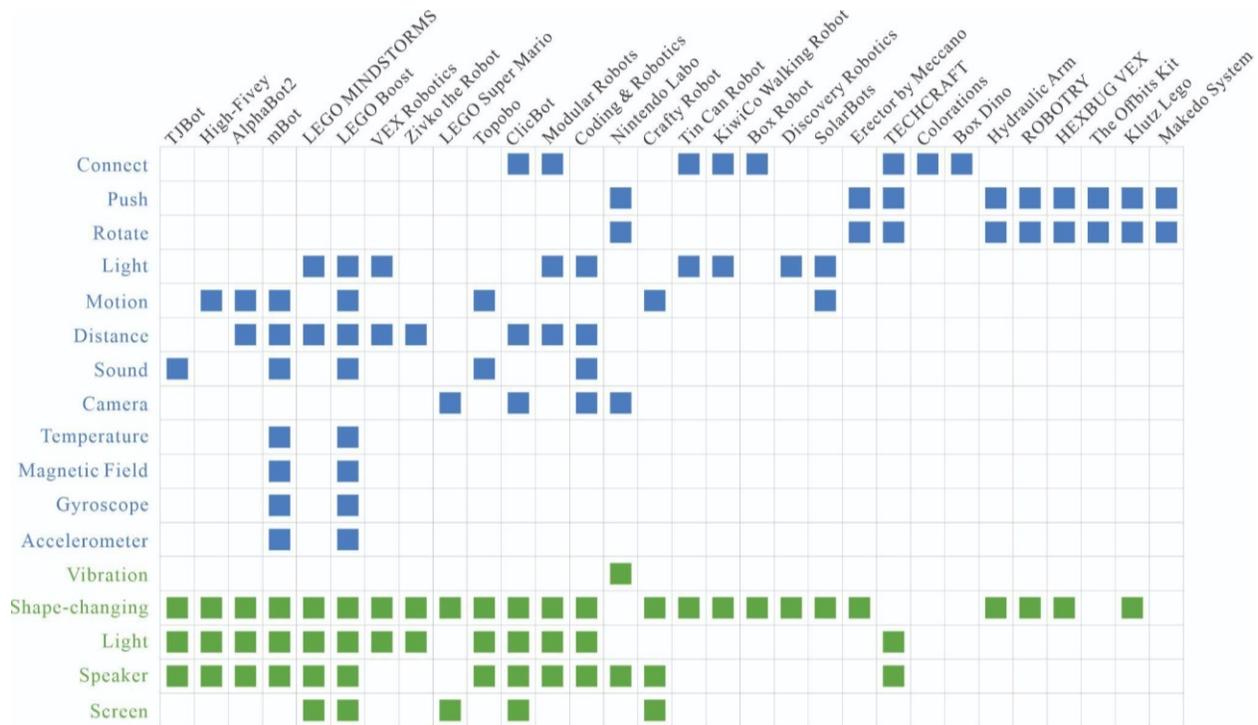

Figure 4: Interaction methods from each robot-building kit

*Input Type 1) Manipulation:* Manipulation refers to direct interaction with the robot. This interaction is usually found in robot-building kits without electronics, or that include only simple circuits without a microcontroller. These kits are simple in structure and less expensive than those with multiple electronics and complex circuitry. The simplest kits are easy to assemble and are often colorful, such as ["Colorations", n.d.]. These kits promote arts and crafts learning, help develop fine motor skills, and encourage curiosity and social development. However, the only "interaction" these robots provide is for the user to **connect** the different parts. Some robot building kits are based on opening and closing circuits [such as Johnco, n.d.], or where the circuit is based on structural connections [such as Schweikardt, 2007]. These kits also use the manipulation of connections as an input. Some robot-building kits focus on the mechanical structure through fine parts to make the whole robot more dynamic, [such as "Nintendo Labo", n.d.]. The mechanical motion of these robots is based on **pushing** and **rotating**.

*Input Type 2) Integrated Sensor:* Integrated Sensor refers to inputs based on electronic components embedded in the robots. Our focus is on the type of interaction, not the type of sensor, so we classify sensors by their ability to detect specific stimuli, rather than their physical design or technical specifications. For example, a phototransistor can detect **light** and also the **motion** of occluders/shadows. So even though there is no controller, such a robot can have input through integrated sensors. Some robot kits use optical encoders to measure positions and motions [such as "LEGO BOOST", n.d.]. Similarly, sensors that detect motion and **distance** are often interchangeable and are common in robot building kits [such as "Zivko the Robot", n.d.]. **Sound** is also a common modality, usually sensed using a microphone [such as "LEGO BOOST", n.d.]. Some more advanced kits use **cameras** to detect different content, including light, color, and barcodes [such as "LEGO Super Mario", n.d.]. Other sensors are only found in advanced robot building kits [such as "MBot | Makeblock", n.d.], including **temperature** sensors, **magnetic field** sensors, **gyroscopes**, and **accelerometers**.

*Output Type 1) Integrated Actuator:* Integrated Actuator refers to outputs based on electronic components embedded in the robots. Almost all robot outputs are of this type, considering that most robots are standalone devices. The most common output is the **shape-changing system** [such as "LEGO BOOST", n.d.], implemented by a variety of motors. Other kits include small motors that output in the form of **vibration** [such as "Nintendo Labo", n.d.]. Another common output is **light**, usually using LEDs or similar optical components. The fourth is the use of **speakers** to provide audio feedback.

*Output Type 2) Separated Actuator:* Separated actuators refer to outputs based on electronic components that are not fully embodied in the robots. They may be connected to the robot by cable or wirelessly, but are not a (structurally or functionally necessary) part of the robot itself. One (and the only) representative actuator is the **screen**. A few robots have a built-in screen, such as ["LEGO Super Mario", n.d., "ClicBot", n.d.], through which they display some interactive information as well as add animation effects to the robot. For other robot-building kits, such as ["LEGO MINDSTORMS", n.d.], the screen gives the user more feedback on the operation than on the interaction. There are also kits involving using a phone or tablet for operation ["LEGO BOOST", n.d., "The Crafty Robot", n.d.], in which case the screen is more a controller than an output device. Except for adding animation effects which are necessary on the robot, all these screen uses do not require the screen itself as part of the robot, so we consider the screen as a separated actuator.

## 4 ELEMENTS OF PAPER-BASED INTERACTIONS

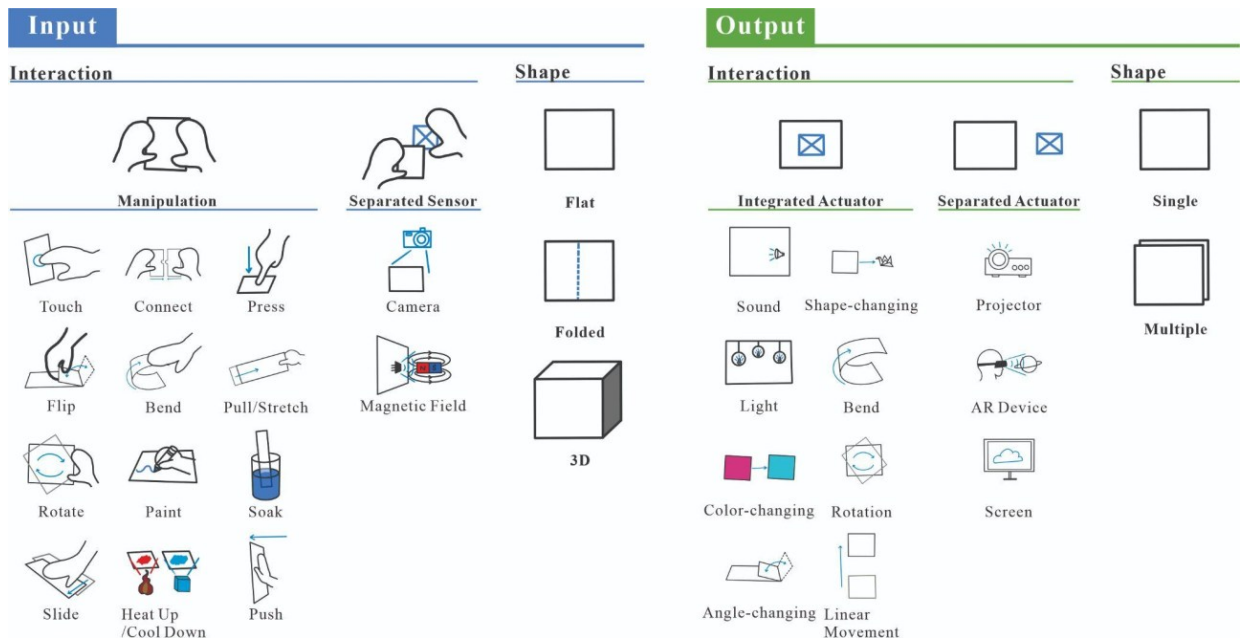

Figure 5: Elements of paper-based interaction designs

In this section, we categorize the interaction designs proposed in paper-based interaction studies (Figure 5). Since the 1990s, many researchers have been exploring paper-based human-computer interaction, a field known as paper computing [Kaplan, 2010]. Research within this field often falls at the intersection of multiple domains, including augmented reality environments, tangible interactions, physical computing, ubiquitous computing, digital art, and more.

Based on the interactive environments, we divide these studies into three groups: 1) interaction based on augmented reality, 2) paper circuits, and 3) movable paper crafts. Inspired by Zhu's taxonomy [2013] we propose another dimension of design space: the *shape* of the paper interface. This dimension spans **flat, folded,** and **3D** paper devices for the input, and **single sheet** and **multiple sheets** of paper for the output devices.

### 4.1 Paper with Augmented Reality (AR)

Paper-based interactions based on augmented reality technologies typically use **manipulation** and **separated sensors** for input, and **separated actuators** for output. Early paper computing focused on the interoperable use of paper and digital documents. These studies aimed to bridge the physical and digital worlds by creating an interactive work environment. Like [Wellner, 1991], many systems used a **camera** to capture the contents of a paper document and the position of the user's finger, and gave feedback using a **projector** to overlay images directly on the same paper document. With the advancement of augmented reality technology, researchers explored designing and developing mixed reality books that use **augmented reality glasses** or **screens** to overlay virtual content onto the pages of physical books [Grasset, 2008, Rajaram, 2022]. These works showed how new interactive technologies could enrich traditional paper-based work. Another series of studies, represented by [Back, 2001, Mackay, 2002, Liao, 2005], were conducted on physical books. These studies used cameras and **magnetic field** sensors to control the audio (**sound**), adding a rich soundtrack to printed graphics and text while maintaining the original appearance of the book. A more recent study uses augmented reality markers (computer vision markers) printed on paper, which turn traditional paper into an interactive interface with a combination of camera and display [Zheng, 2020, Bae, 2021, Bae, 2022]. Across all these studies, an important insight of paper-based interaction is to maintain the properties of the paper and the ways people traditionally interact with them. Users can interact with those through conventional methods of manipulation, such as **touching**, **flipping**, **rotating**, and **sliding**.

Regarding the shape of the paper, the input of most systems [Wellner, 1991, Grasset, 2008, Back, 2001, Mackay, 2002, Liao, 2005] is based on **flat** paper except for [Zheng, 2020] which is based on **folded** paper and [Bae, 2022] which is based on **3D** paper. Systems that use projectors output on the **single sheet** of paper; unlike systems that use screens.

### 4.2 Paper Circuit

Most paper circuits use **manipulation** as input, and **integrated actuators** for outputs, with all outputs directly on the paper. Starting with pulp-based computing, researchers investigated embedded circuits. Some start from the process of making paper by hand and embedding conductive materials or electronic components into the paper "sandwich" [Coelho, 2009, Knouf, 2017]. Other studies proposed different ways of making paper circuits, including the use of circuit stickers [Hodges, 2014, "Chibitronics", n.d.] and weaving techniques [Kato, 2019]. In these systems, common inputs include **connection** and **bending**, and common outputs include embedded LEDs (**lights**). The inputs of these interactions are based on **3D** structures, while the outputs are based on **multiple sheets** of paper.

The other direction of exploration focuses on the surface of the paper. By adding different coatings and pasting different electronic components, [Buechley, 2009] introduces interaction using **painting** as input. Further explorations of material properties include paper's thinness and softness. Researchers further explored the coating of the paper. These studies discussed the fabrication of paper-based interaction with mechanical processing equipment instead of traditional handcrafting. Using color-changing paint, researchers created paper computing artworks that would make these graphic arts pieces more interactive [Kato, 2019]. In this interaction, the **color-changing** was triggered by **heating up** and **cooling down** the paper. The inputs of these interactions are based on **flat** paper, while the outputs are based on a **single sheet** of paper.

As the forms of paper computing became diverse, researchers began to explore the aesthetics of this field. With the Electronic Popables proposed, the combination of interactive books and electronic devices was also explored [Qi, 2010]. In this design, the input modalities include rotation, touching, **pressing**, **pulling**, and painting, with the light output through the embedded LEDs. The inputs of these interactions are based on **folded** paper, while the outputs are based on **multiple sheets** of paper.

Techniques for making circuits via inkjet printers have also been widely discussed. They are commonly used to make paper-based sensors and printed circuit boards [Oh, 2018, Gong, 2014, Landers, 2022]. Paper circuits can be used to make loudspeakers (**sound**) without permanent magnets via electrostatic speaker technology [Kato, 2022]. The inputs presented in these systems include connection, temperature-changing, bending, pressing, touching, and **soaking**; the outputs include lights, color-changing, and sound. In addition to adding conductive materials to paper, researchers have also explored working on coated paper. In Sensing Kirigami, researchers enabled the paper to sense bending, **pushing**, and **stretching**, by making different cuts to carbon-coated paper [Zheng, 2019]. Along with the exploration of materials, these studies also emphasize the importance of aesthetics in paper computing, opening up more possibilities for the fabrication of paper-based interaction. The inputs of these interactions are based on **3D** paper structures; while the output of [Kato, 2022] is based on a **single sheet** of paper and [Oh, 2018] is based on **multiple sheets** of paper. Most systems from this series are served as input devices only.

### 4.3 Moveable Paper Crafts

Interactions in moveable paper crafts are more focused on outputs, including both **integrated actuators** and **separated actuators**. [Wrensch, 1998, Saul, 2010, Zhu, 2013] investigated shape memory alloy (SMA) embedded in paper objects. Multiple patterned SMA wires can be straightened or bent upon receipt of a signal, thus triggering **angle-changing**. These movements can occur on a whole paper structure (**single sheet**) [Wrensch, 1998], between two paper structures connected by SMA wires (**multiple sheets**) [Saul, 2010, Zhu, 2013], or on a whole sheet of folded paper (**single sheet**) [Qi, 2012]. Due to the limitations of SMA materials in handling, other deformation materials have also been investigated. Some researchers have emphasized the importance of making deformable paper crafts at home, and suggested using microwaves to heat paper for **shape-changing** [Yasu, 2012]. To that end, the combination of paper and plastic was also explored. By using 3D printers to add Polylactide (PLA) to the surface of the paper, these novel and easy-to-manufacture paper actuators offer reversible deformation (**bending**) [Wang, 2018]. Using the swelling and shrinking (shape-changing) of paper, researchers also printed wax on paper to create a water-driven paper robot [Ryu, 2020]. The outputs of these systems are all based on a **single sheet** of paper.

In addition to these movements with embedded materials, paper movements using separate actuators have also been explored. [Oh, 2018] presented different motor-based paper movements, including **rotation** and **linear movement**. This system demonstrated the great potential of this technology for education, machines, toys, and dynamic artwork. By adding magnets to the paper and placing it on a base with motors, researchers have also developed low-cost toolkits that allow non-technical designers to see shape-changing more intuitively [Yehoshua, 2021]. These systems are output on **multiple sheets** of paper. These explorations often emphasize low cost, creativity, and customization. They connect the two fields of handcraft and computing, introducing the concept of computationally-enhanced craft.

## 5 DESIGN SPACE OF PAPER ROBOTS

Building on the relevant work of robot-building kits and paper-based interactions, this section proposes the design space of paper robots (review Figure 1 for a visual summary of the design space dimensions). For the interface shapes, we used the same dimensions of paper-based interaction design space – flat, folded, and 3D paper structures for input, and single and multiple sheets of paper for output. For each individual interaction approach, we explain it with illustrations and text in the following tables (Table 1 and Table 2). In particular, when introducing designs from the robot-building kits into this design space, we made some adjustments to their type and approach.

This design space covers the interactions that appear in the paper-based interaction designs and the robot-building kits, helping us to understand the capabilities and limitations of the prior work. We first noticed that prior paper-based interaction designs did not cover all interactions we can find in the robot-building kits. Most paper-based interaction designs have focused on manipulation of the paper itself, and less on sensing with embedded electronics. For example, distance sensing with ultrasonic sensors, which is common in robot-building kits, has not yet been used in paper-based interaction design. In addition to compiling the interactions from the robot-building kits and paper-based interaction design, we propose additional interactions based on our own experience on paper crafts, including inputs such as twist, blow, stack, cut, moisture, AR glasses, and motion capture system, and outputs such as twist, connection, separation, length-changing, and shadow. The categories of these interactions are briefly illustrated in the figure below (Figure 6).

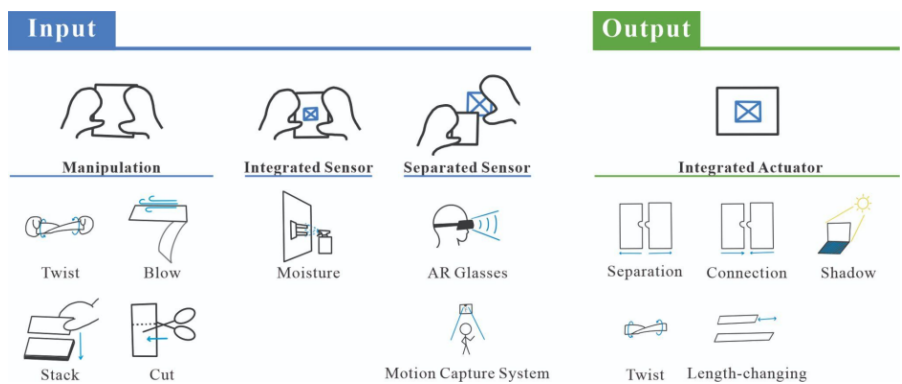

Figure 6: Additional interaction methods

With this design space being proposed, we can perceive these missing interactions and propose new designs. Most of the robot-building kits use plastic or metal parts, which are expensive and not easily customized. Given the current cardboard robot-building kits available, it is feasible to use paper for structural construction. And with this design space, we see the possibility of using paper instead of traditional plastic parts to complete the interaction, which can further reduce the cost of the robot-building kits, and make them easy to customize.

Table 1: Interaction modalities as inputs

| Manipulation | | | |
|---|---|---|---|
| 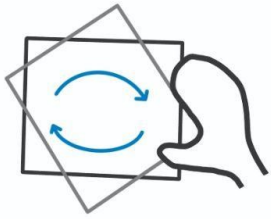 **Rotate:** Input by rotating the paper | 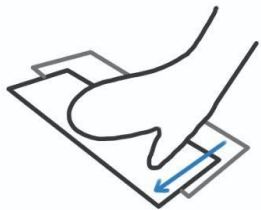 **Slide:** Input by moving the paper linearly | 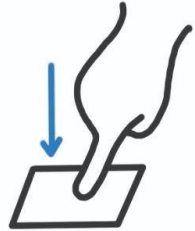 **Press:** Input by pressing on the paper | 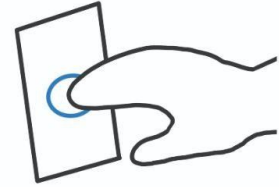 **Touch:** Input by touching the surface of the paper |
| 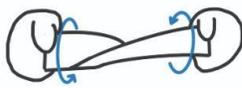 **Twist:** Input by twisting the paper | 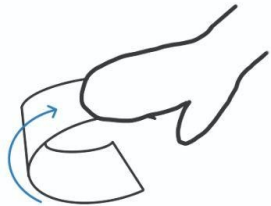 **Bend:** Input by bending the paper | 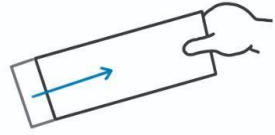 **Pull/Stretch:** Input by pulling or stretching the paper | 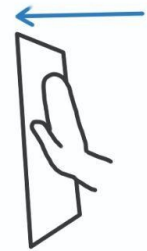 **Push:** Input by pushing the paper |
| 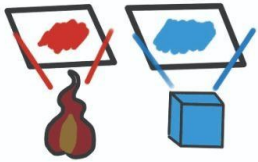 **Heat Up/Cool Down:** Input by changing the temperature of the paper | 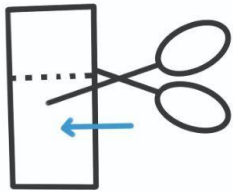 **Cut:** Input by cutting the paper | 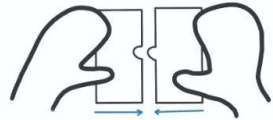 **Connect:** Input by connecting separate sheets of paper | 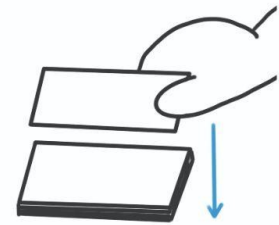 **Stack:** Input by stacking separate sheets of paper |
| 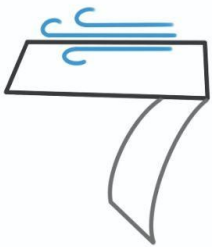 **Blow:** Input by blowing air on or into the paper | 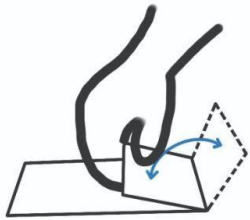 **Flip:** Input by flipping the paper | 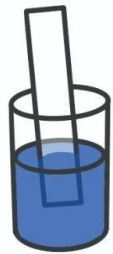 **Soak:** Input by soaking the paper in water | 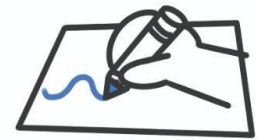 **Paint:** Input by drawing on paper |

Table 1 (continued): Interaction modalities as inputs

## Integrated Sensor

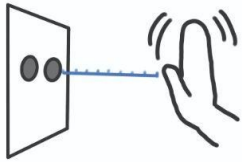

**Distance:** Input through distance sensors embedded in the paper

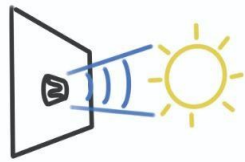

**Light:** Input through photoelectric sensors embedded in the paper

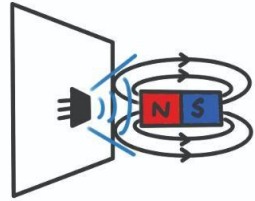

**Magnetic Field:** Input through magnetic field sensors embedded in the paper

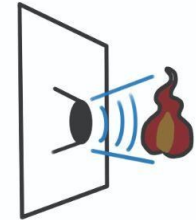

**Temperature:** Input through temperature sensors embedded in the paper

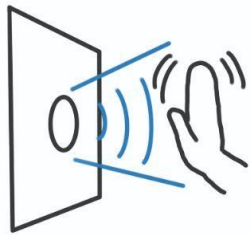

**Motion:** Input through motion sensors embedded in the paper

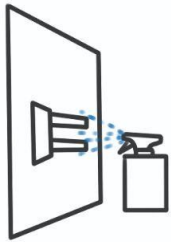

**Moisture:** Input through moisture sensors embedded in the paper

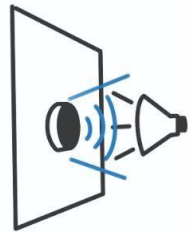

**Sound:** Input through the microphone embedded in the paper

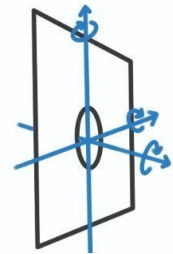

**Gyroscope:** Input through angular velocity sensors embedded in the paper

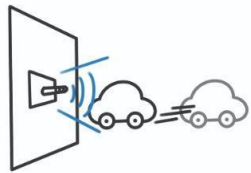

**Accelerometer:** Input through acceleration sensors embedded in the paper

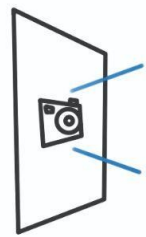

**Camera:** Input through cameras embedded in the paper

## Separated Sensor

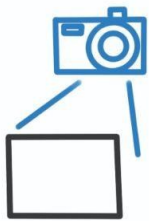

**Camera:** Input through separate cameras

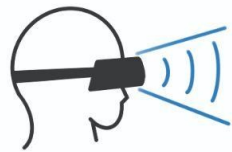

**AR Glasses:** Input through separate AR glasses

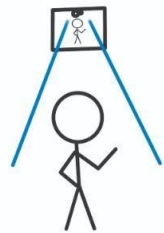

**Motion Capture System:** Input through separate motion capture systems

Table 2: Interaction modalities as outputs

### Integrated Actuator

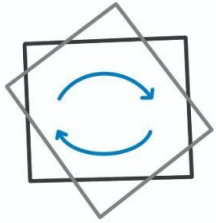
**Rotation:** Paper being rotated

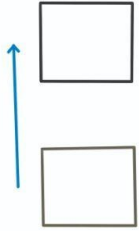
**Linear Movement:** Paper being moved linearly

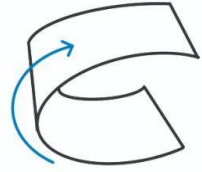
**Bend:** Paper being bent

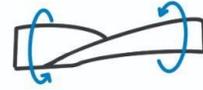
**Twist:** Paper being twisted

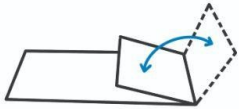
**Angle-changing:** Change in the angle of the paper

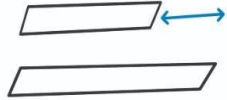
**Length-changing:** Change in the length of the paper

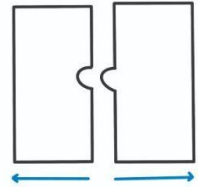
**Separation:** Paper being separated

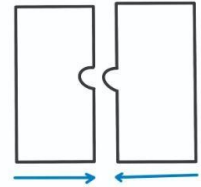
**Connection:** Paper being connected

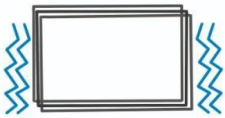
**Vibration:** Paper being vibrated

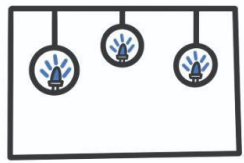
**Light:** Paper (or the illuminated object on it) being lit up

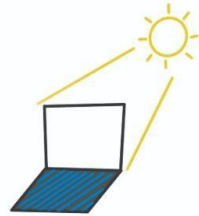
**Shadow:** Shadows being formed

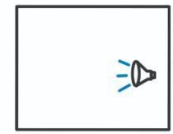
**Sound:** Sound being made

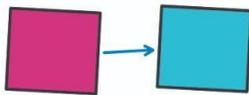
**Color-changing:** Change in the color of the paper

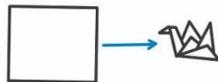
**Shape-changing:** Change in the shape of the paper

Table 2 (continued): Interaction modalities as outputs

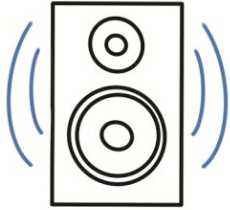
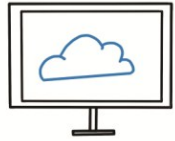
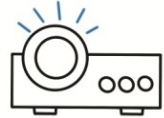
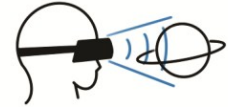

**Speaker:** Sound coming from separate speakers

**Screen:** Image or text display through separate screens

**Projector:** Image or text display through separate projectors

**AR device:** Image or text display through separate AR devices

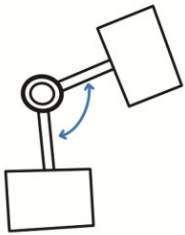

**Shape-changing system:** Change in the shape of the paper with the separated system

## 6 DESIGN EXAMPLES OF PAPER ROBOT-BUILDING KITS

Robot-building kits with paper-based interactions provide a balance between the cost, customization, and functionality. With this vision, we designed three paper circuit building kits and two paper robot-building kits. The paper circuit building kits focus on visual feedback from the circuits, while the paper robot-building kits provide more dynamic interactions and enable end-users to customize these units in terms of both digital creation and physical fabrication. In the following, we present the designs of these building kits and evaluate them based on our proposed design space.

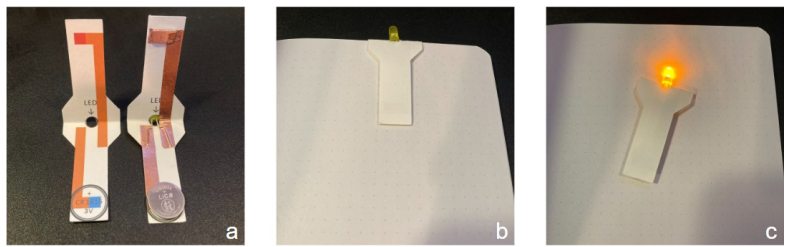

Figure 7: *Bookmark Light Kit* (a: printed pattern (left) and built bookmark (right); b: the light is off when the bookmark is on the page; c: and lights up when it's removed from the page)

The *Bookmark Light Kit* (Figure 7) provides materials to make a bookmark that lights up. After completing the circuit as illustrated, the bookmark attaches to the pages of a book like a normal magnetic bookmark, and its circuit will automatically close when it is removed thus becoming a small flashlight. This kit uses **manipulation-connection** and **folded** paper structures for the input, and **integrated actuator-light** with **multiple sheets** of paper for the output. Although it is folded from a single sheet, its circuit is divided into two parts, one on each side of the folding line. In this case we considered it as multiple sheets.

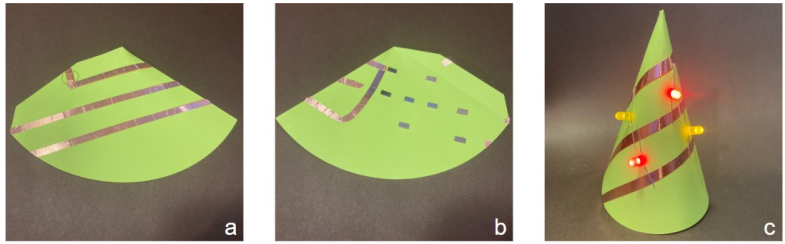

Figure 8: *Christmas Tree Kit* (a: front side (outside) of the paper; b: back side (inside) of the paper; c: built Christmas tree (bent and standing))

The *Christmas Tree Kit* (Figure 8) is a Christmas card that can be mailed to others. After applying the copper tape and magnets to the cardstock, users bend it into a cone and then use the attraction of the magnets to hang LEDs on it. Its circuit is more complex than the previous kit, requiring copper tape on both sides of the paper. This kit also uses **manipulation-connection** as an input, but it is based on a **3D** paper structure, as the completion of the circuit relies on the 3D structure after bending. It uses **integrated actuator-light** as its output. Although its circuits are on the front and back sides of the paper, these circuits cannot be physically separated as in the previous set, in which case we consider the output to be based on a **single sheet** of paper.

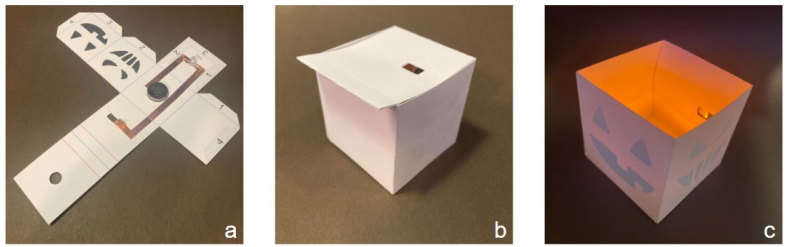

Figure 9: *Halloween Lantern Kit* (a: half-built lantern(flat); b: built lantern (turned off); c: built lantern (turned on))

The third kit we designed omits the magnets and uses a paper structure to open and close the circuit. The *Halloween Lantern Kit* (Figure 9) provides a pattern of a box lantern with a movable panel that acts as an on/off switch. Once built, the lantern looks like a white cube; upon pressing the movable panel, the LED in the middle of the lantern will light up and the design printed inside the lantern will be displayed. Its inputs are **manipulation-pressing** and **manipulation-pulling** based on the **3D** paper structure; its outputs are **integrated actuator-lights** and **integrated actuator-shadow** based on the **single sheet** of paper structure.

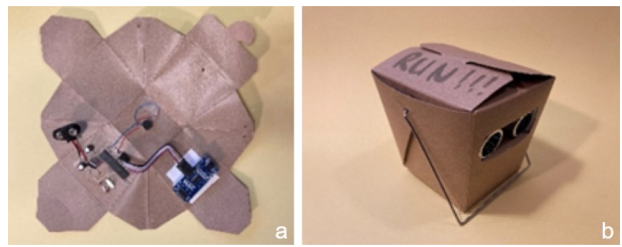

Figure 10: The *Escaping Takeout Box* (a: flat; b: folded)

The *Escaping Takeout Box* (Figure 10) is a stand-alone paper robot. This paper robot-building kit includes controller circuits, a battery, two vibration motors, an ultrasonic sensor, and a piece of cardstock that folds into the shape of a takeout box. After it is built, the pre-burned program in the controller enables it to detect nearby objects and "escapes" when something (such as a human hand) approaches it. End-users can customize this robot in both physical fabrication and digital creation. This building kit comes with a pre-cut cardstock piece that can be trimmed to any new pattern for physically customizing the robot's structure. The ATMEGA328P microcontroller used in this robot allows for smooth programming and customization by end-users. It is compatible with the Arduino Uno platform and can read and write programs directly, without any additional equipment or chip burners. This robot uses **integrated sensor-distance** for input with a **3D** paper structure, and the output is based on the **single sheet** of paper, moving by **integrated actuator-vibration** that changes its center of gravity.

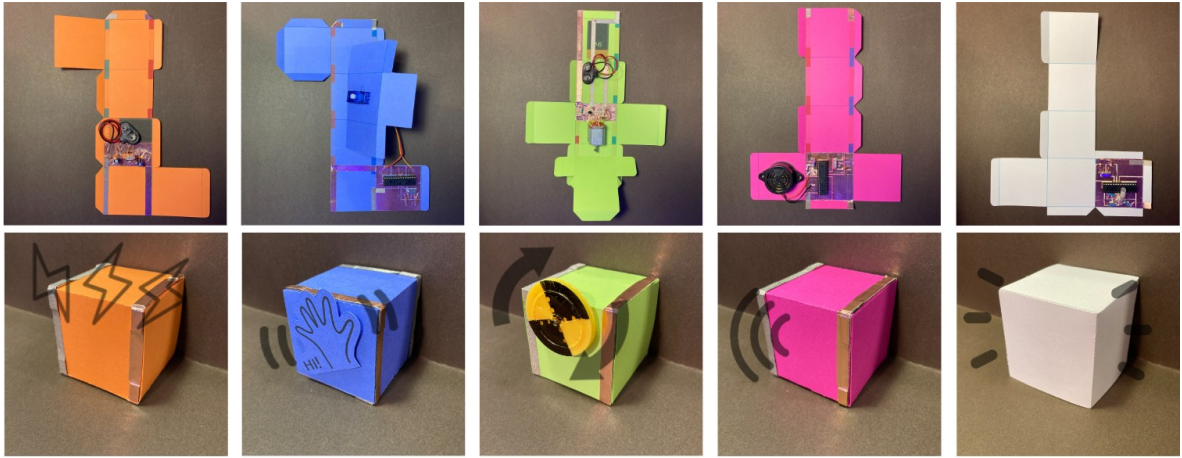

Figure 11: *Paper Box Robot Kit* (from left to right: power box, waving box, spinning box, speaker box, and light box)

The *Paper Box Robot Kit* (Figure 11) [Yang, 2024] uses a modular design that was inspired by modular robots like Topobo [Raffle, 2004] and Cubelets [Schweikardt, 2007]. This robot includes five modules, which are the *power box*, the *waving box*, the *spinning box*, the *speaker box*, and the *light box*. Except for the *power* and *spinning* boxes, the other three boxes have independent controller circuits. Similar to the *Escaping Takeout Box*, these controllers offer the capability of being reprogrammed by end-users, and the use of IC sockets makes it easier for users to replace these microcontrollers. Each of these boxes could be considered as an individual paper circuit building kit, and they combine to form a paper robot-building kit. We color-coded the different boxes in order to better distinguish them as they are similar in appearance.

As an entire paper robot-building kit, the Paper Box Robot Kit features inputs that are based on **3D** paper structure with **manipulation-connection**. The *power box* (orange) contains a simple circuit with a 9v battery, two capacitors, and a voltage regulator. The conductive fabric tape on the outside of the box passes through the four sides of the box and then goes inside the box to connect with the two terminals of the circuit output.

The *waving box* (blue) includes the controller circuit and a servo motor held in the middle by a paper structure, and its output is based on **multiple sheets** of paper with **integrated actuator-rotation**. The motor connects to the piece of paper on the outside of the box by magnetic coupling ["Engineering360", n.d.], to transmit torque without touching. Magnetic couplings allow the robot to obtain more mechanical motion while being easy to assemble.

The *spinning box* (green) uses a DC motor, and as it requires more power to operate, we added an extra battery, a diode, and a transistor to the circuit. A small opening on the box allows the motor shaft to go straight through the box to the outside, so its output is based on a **single sheet** of paper with the **separated actuator-shape-changing system**. It differs t from the *waving box* in that its output is the motor itself, not on the paper. If we connect the motor's shaft to another box, then its output will be based on **multiple sheets** of paper. Due to the low torque of the motor in the *spinning box*, magnetic coupling is not suitable.

The *speaker box* (red) includes the control circuit, an additional resistor, and a buzzer. It uses a **single sheet** of paper with the **integrated actuator-sound** as output.

The *light box* (white) includes the control circuit, three resistors, and an RGB LED. This box also comes with an outer shell, the inside of which can be printed or painted with different patterns, and the patterns become visible when the shell is set on the lighted box. In addition to using manipulation-connection, the combination of the shell and the *light box* also use **manipulation-stacking** as input. The output of the light box is based on **integrated actuators-lights** with the **single sheet** structure when used alone, and **integrated actuator-shadow** with **multiple sheets** when used with a shell.

**7 DISCUSSION AND CONCLUSION**

Based on a review of 60 relevant works, we proposed a design space for paper robots and presented five building kits that were designed using elements from this design space. Our workshops confirmed the public's interest in building robots out of paper. Our design space can help researchers and designers better understand the state of the art in paper-based interaction design, and we can address the gaps between paper-based interactions and robot-building kit designs by filling in the empty slots of the design space. The building kits presented in this paper serve as examples of how elements in the design space could be combined and applied to paper robots.

The design space reveals that most current paper-based interactions consist of manipulation as inputs with integrated actuators for outputs. Based on these interactions that have already been developed, we can build paper robots with different combinations of them. With the design space, there are also opportunities to build paper robots by combining more types of inputs and outputs. However, for paper-based interactions with separate sensors and separate actuators, how to combine these electronic components with paper remains a key issue for researchers.

The current design space for paper robots has certain limitations, such as the lack of analysis on controllers and other logic units. In our future work, more dimensions of this design space will be developed to better support complex interaction designs. Another question to ponder is whether there should be a limit to one input and one output in a paper robot. Although the *Paper Box Robot Kit* provided multiple output units, participants in the workshop preferred to connect one box at a time to the power unit; however, for the *Escaping Takeout Box*, participants did expect more feedback than just vibration/movement after a certain amount of time spent with it. In addition, the exploration of the appearance of paper robots should also be taken into consideration, in other words, appearance and size should also be included in the design space in the future.

Another dimension that needs to be developed is the fabrication/building technique, and this is also an issue that must be addressed. In discussing the accessibility of paper robot-building kits, we would like the fabrication techniques used to be household-accessible and not requiring any special environment or equipment. Accessibility was the main reason we chose the paper structures and circuit designs that are full of straight lines. While using copper tape is an affordable option, during the workshop, we noted that some participants were unable to use copper tape proficiently, especially children and others with fine motor impairments. The use of conductive paint, on the other hand, would have been more costly, which was not in line with the original intent of building robots with paper. As the designs of paper robots become increasingly complex, it is critical that we not only consider their interaction elements, but also investigate the methods used to fabricate them. While copper tape is currently sufficient for building relatively simple circuits, we must explore alternative solutions to make the fabrication of paper circuits easier and more feasible for future designs. By bringing fabrication methods into the design space, we can ensure that paper robots are accessible and feasible.

Paper robot-building kits achieve a good balance between affordability, functionality and customization options. Building robots out of paper allows for more creative expressions. People can use different colors, patterns and shapes to make the robot the way they like without worrying about the cost or recycling of the materials. Looking at the current market, the maker community, and the research community, there are already some robots made using cardboard. However, it wouldn't make the most sense to just replace the plastic material of a traditional robot with cardboard and then use plastic parts or metal screws to connect them while keeping the traditional circuitry. We need paper robots that highlight the properties of the paper itself, and by integrating paper-based interaction design, we can achieve this goal and create real paper robots. For future work, we should further explore the paper robot design space and make better use of paper-based interactions for different robot designs.

In this paper, we present the design space for paper robots, which brings together insights from traditional robot-building kits and paper-based interaction studies. Our aim is to create a comprehensive framework to explore the potential of paper robots and to foster creativity in this emerging field. We hope that our work will provide guidance for future investigations and stimulate future studies of paper robots.

# REFERENCE

4M Kidzlabs Mega Hydraulic Arm Robotic Science Kit. Walmart.com. https://www.walmart.com/ip/4M-Kidzlabs-Mega-Hydraulic-Arm-Robotic-Science-Kit/396388274

AlphaBot2 robot building kit for BBC micro:bit (no micro:bit). botnroll.com. https://www.botnroll.com/en/mounting-kits/3523-alphabot2-robot-building-kit-for-bbc-micro-bit-no-micro-bit.html

Chibitronics. Chibitronics. https://chibitronics.com/

Colorations® Create Your Own Robot - Kit for 12. Discount School Supply. https://www.discountschoolsupply.com/arts-crafts/arts-crafts-kits/craft-kits-projects/colorations-create-your-own-robot---kit-for-12/p/33667

Discovery Robotics. Walmart.com. https://www.walmart.com/ip/Discovery-Robotics/341579728

Magnetic Couplings Selection Guide: Types, Features, Applications | Engineering360. https://www.globalspec.com/learnmore/motion_controls/power_transmission_mechanical/magnetic_couplings

Amazon.com: Erector by Meccano Super Construction 25-In-1 Motorized Building Set, Steam Education Toy, 638 Parts, For Ages 10+ : Toys & Games. https://www.amazon.com/Meccano-Construction-Motorized-Building-Education/dp/B000GOF5S2

"High-Fivey" the Cardboard Micro:bit Robot : 18 Steps (with Pictures). Instructables. https://www.instructables.com/High-Fivey-the-Cardboard-Microbit-Robot/

Amazon.com: Klutz Lego Gear Bots Science/STEM Activity Kit : Toys & Games. https://www.amazon.com/Klutz-Lego-Gear-Bots/dp/1338603450

LEGO® BOOST | Official LEGO® Shop US. https://www.lego.com/en-us/themes/boost/about

LEGO® MINDSTORMS® | Invent a Robot | Official LEGO® Shop US. https://www.lego.com/en-us/themes/mindstorms

Take brick building to a new level with LEGO® Super Mario™. https://www.lego.com/en-us/themes/super-mario/about

Makedo. https://www.make.do/

mBot Neo STEM Programmable Robotics Kit | Makeblock. https://store.makeblock.com/products/diy-coding-robot-kits-mbot-neo?currency=USD&variant=42817922891992&utm_medium=cpc&utm_source=google&utm_campaign=Google%20Shopping&gclid=CjwKCAiA_vKeBhAdEiwAFb_nraXwT2YLaugSs8S8sUFNKRbd7elE42jax0VgWbD_6yaldKn56vtSaBoCWzQQAvD_BwE

Nintendo Labo Toy-Con 01 Variety Kit | Nintendo Switch | Nintendo. Nintendo Labo Toy-Con 01 Variety Kit | Nintendo Switch | Nintendo. https://www.nintendo.com/sg/switch/adfu/index.html

Amazon.com: Thames & Kosmos SolarBots: 8-in-1 Solar Robot STEM Experiment Kit | Build 8 Cool Solar-Powered Robots in Minutes | No Batteries Required | Learn About Solar Energy & Technology | Solar Panel Included : Toys & Games. https://www.amazon.com/Thames-Kosmos-SolarBots-Experiment-Solar-Powered/dp/B085P361MQ

Kids First Coding & Robotics Screen-free Coding Kit & Lessons, K to 2 – Thames & Kosmos. https://store.thamesandkosmos.com/products/coding-and-robotics

The Crafty Robot. The Crafty Robot. https://thecraftyrobot.net/

Jumbo Kit > TheOffbits. TheOffbits. https://theoffbits.com/product/offbits-multi-kit-jumbo-kit

VEX Robotics | HEXBUG. https://www.hexbug.com/vex

Walking Robot. KiwiCo. https://www.kiwico.com/us/store/dp/walking-robot-project-kit/1986

Zivko the Robot. https://shop.elenco.com/consumers/zivko-the-robot.html

로보트리(Robotry). https://robotry.co.kr/

Maribeth Back, Jonathan Cohen, Rich Gold, Steve Harrison, and Scott Minneman. 2001. Listen reader: an electronically augmented paper-based book. In Proceedings of the SIGCHI Conference on Human Factors in Computing Systems (CHI '01), 23–29. https://doi.org/10.1145/365024.365031

S. Sandra Bae, Rishi Vanukuru, Ruhan Yang, Peter Gyory, Ran Zhou, Ellen Yi-Luen Do, and Danielle Albers Szafir. 2022. Cultivating Visualization Literacy for Children Through Curiosity and Play. https://doi.org/10.48550/arXiv.2208.05015

Sandra Bae, Ruhan Yang, Peter Gyory, Julia Uhr, Danielle Albers Szafir, and Ellen Yi-Luen Do. 2021. Touching Information with DIY Paper Charts & AR Markers.

In Interaction Design and Children (IDC '21), 433–438. https://doi.org/10.1145/3459990.3465191

Leah Buechley, Sue Hendrix, and Mike Eisenberg. 2009. Paints, paper, and programs: first steps toward the computational sketchbook. In Proceedings of the 3rd International Conference on Tangible and Embedded Interaction (TEI '09), 9–12. https://doi.org/10.1145/1517664.1517670

Marcelo Coelho, Lyndl Hall, Joanna Berzowska, and Pattie Maes. 2009. Pulp-based computing: a framework for building computers out of paper. In CHI '09 Extended Abstracts on Human Factors in Computing Systems (CHI EA '09), 3527–3528. https://doi.org/10.1145/1520340.1520525

Victor C. Dibia, Maryam Ashoori, Aaron Cox, and Justin D. Weisz. 2017. TJBot: An Open Source DIY Cardboard Robot for Programming Cognitive Systems. In Proceedings of the 2017 CHI Conference Extended Abstracts on Human Factors in Computing Systems (CHI EA '17), 381–384. https://doi.org/10.1145/3027063.3052965

Susan Freinkel. 2011. Plastic: A Toxic Love Story. Text Publishing Company.

Nan-Wei Gong, Jürgen Steimle, Simon Olberding, Steve Hodges, Nicholas Edward Gillian, Yoshihiro Kawahara, and Joseph A. Paradiso. 2014. PrintSense: a versatile sensing technique to support multimodal flexible surface interaction. In Proceedings of the SIGCHI Conference on Human Factors in Computing Systems (CHI '14), 1407–1410. https://doi.org/10.1145/2556288.2557173

Raphaël Grasset, Andreas Dünser, and Mark Billinghurst. 2008. Edutainment with a mixed reality book: a visually augmented illustrative childrens' book. In Proceedings of the 2008 International Conference on Advances in Computer Entertainment Technology (ACE '08), 292–295. https://doi.org/10.1145/1501750.1501819

Steve Hodges, Nicolas Villar, Nicholas Chen, Tushar Chugh, Jie Qi, Diana Nowacka, and Yoshihiro Kawahara. 2014. Circuit stickers: peel-and-stick construction of interactive electronic prototypes. In Proceedings of the SIGCHI Conference on Human Factors in Computing Systems (CHI '14), 1743–1746. https://doi.org/10.1145/2556288.2557150

Johnco. 4M - KidzRobotix - Tin Can Robot. Johnco. http://www.johncoproductions.com/products/4m-kidzrobotix-tin-can-robot

Johnco. 4M - Sci:Bits - Box Robot. Johnco. http://www.johncoproductions.com/products/4m-sci-bits-box-robot

Johnco. 4M - Techcraft - Paper Circuit Science. Johnco. http://www.johncoproductions.com/products/4m-techcraft-sound-light-kit

Frédéric Kaplan and Patrick Jermann. 2010. PaperComp 2010: first international workshop on paper computing. In Proceedings of the 12th ACM international conference adjunct papers on Ubiquitous computing - Adjunct (UbiComp '10 Adjunct), 507–510. https://doi.org/10.1145/1864431.1864500

Kunihiro Kato, Kaori Ikematsu, Yuki Igarashi, and Yoshihiro Kawahara. 2022. Paper-Woven Circuits: Fabrication Approach for Papercraft-based Electronic Devices. In Sixteenth International Conference on Tangible, Embedded, and Embodied Interaction (TEI '22), 1–11. https://doi.org/10.1145/3490149.3502253

Kunihiro Kato, Kazuya Saito, and Yoshihiro Kawahara. 2019. OrigamiSpeaker: Handcrafted Paper Speaker with Silver Nano-Particle Ink. In Extended Abstracts of the 2019 CHI Conference on Human Factors in Computing Systems (CHI EA '19), 1–6. https://doi.org/10.1145/3290607.3312872

Nicholas A. Knouf. 2017. Felted Paper Circuits Using Joomchi. In Proceedings of the Eleventh International Conference on Tangible, Embedded, and Embodied Interaction (TEI '17), 443–450. https://doi.org/10.1145/3024969.3025071

Mya Landers, Anwar Elhadad, Maryam Rezaie, and Seokheun Choi. 2022. Integrated Papertronic Techniques: Highly Customizable Resistor, Supercapacitor, and Transistor Circuitry on a Single Sheet of Paper. ACS Applied Materials & Interfaces 14, 40: 45658–45668. https://doi.org/10.1021/acsami.2c13503

Chunyuan Liao, François Guimbretière, and Ken Hinckley. 2005. PapierCraft: a command system for interactive paper. In Proceedings of the 18th annual ACM symposium on User interface software and technology (UIST '05), 241–244. https://doi.org/10.1145/1095034.1095074

Wendy E. Mackay, Guillaume Pothier, Catherine Letondal, Kaare Bøegh, and Hans Erik Sørensen. 2002. The missing link: augmenting biology laboratory notebooks. In Proceedings of the 15th annual ACM symposium on User interface software and technology (UIST '02), 41–50. https://doi.org/10.1145/571985.571992

Hyunjoo Oh, Sherry Hsi, Michael Eisenberg, and Mark D. Gross. 2018. Paper mechatronics: present and future. In Proceedings of the 17th ACM Conference on Interaction Design and Children (IDC '18), 389–395. https://doi.org/10.1145/3202185.3202761

Hyunjoo Oh, Tung D. Ta, Ryo Suzuki, Mark D. Gross, Yoshihiro Kawahara, and Lining Yao. 2018. PEP (3D Printed Electronic Papercrafts): An Integrated Approach for 3D Sculpting Paper-Based Electronic Devices. In Proceedings of the 2018 CHI Conference on Human Factors in Computing Systems (CHI '18), 1–12. https://doi.org/10.1145/3173574.3174015

Jie Qi and Leah Buechley. 2010. Electronic popables: exploring paper-based computing through an interactive pop-up book. In Proceedings of the fourth international conference on Tangible, embedded, and embodied interaction (TEI '10), 121–128. https://doi.org/10.1145/1709886.1709909

Jie Qi and Leah Buechley. 2012. Animating paper using shape memory alloys. In Proceedings of the SIGCHI Conference on Human Factors in Computing Systems (CHI '12), 749–752. https://doi.org/10.1145/2207676.2207783

Hayes Solos Raffle, Amanda J. Parkes, and Hiroshi Ishii. 2004. Topobo: a constructive assembly system with kinetic memory. In Proceedings of the 2004 conference on Human factors in computing systems - CHI '04, 647–654. https://doi.org/10.1145/985692.985774

Shwetha Rajaram and Michael Nebeling. 2022. Paper Trail: An Immersive Authoring System for Augmented Reality Instructional Experiences. In CHI Conference on Human Factors in Computing Systems (CHI '22), 1–16. https://doi.org/10.1145/3491102.3517486

M. Resnick, F. Martin, R. Sargent, and B. Silverman. 1996. Programmable Bricks: Toys to think with. IBM Systems Journal 35, 3.4: 443–452. https://doi.org/10.1147/sj.353.0443

Jihyun Ryu, Maedeh Mohammadifar, Mehdi Tahernia, Ha-ill Chun, Yang Gao, and Seokheun Choi. 2020. Paper Robotics: Self-Folding, Gripping, and Locomotion. Advanced Materials Technologies 5, 4: 1901054. https://doi.org/10.1002/admt.201901054

Greg Saul, Cheng Xu, and Mark D. Gross. 2010. Interactive paper devices: end-user design & fabrication. In Proceedings of the fourth international conference on Tangible, embedded, and embodied interaction (TEI '10), 205–212. https://doi.org/10.1145/1709886.1709924

Eric Schweikardt. 2007. Modular robotics as tools for design. In Proceedings of the 6th ACM SIGCHI conference on Creativity & cognition (C&C '07), 298. https://doi.org/10.1145/1254960.1255034

Kohei Tsuji and Akira Wakita. 2011. Anabiosis: an interactive pictorial art based on polychrome paper computing. In Proceedings of the 8th International Conference on Advances in Computer Entertainment Technology - ACE '11, 1. https://doi.org/10.1145/2071423.2071521

Guanyun Wang, Tingyu Cheng, Youngwook Do, Humphrey Yang, Ye Tao, Jianzhe Gu, Byoungkwon An, and Lining Yao. 2018. Printed Paper Actuator: A Low-cost Reversible Actuation and Sensing Method for Shape Changing Interfaces. In Proceedings of the 2018 CHI Conference on Human Factors in Computing Systems (CHI '18), 1–12. https://doi.org/10.1145/3173574.3174143

Pierre Wellner. 1991. The DigitalDesk calculator: tangible manipulation on a desk top display. In Proceedings of the 4th annual ACM symposium on User interface software and technology (UIST '91), 27–33. https://doi.org/10.1145/120782.120785

Thomas Wrensch and Michael Eisenberg. 1998. The programmable hinge: toward computationally enhanced crafts. In Proceedings of the 11th annual ACM symposium on User interface software and technology (UIST '98), 89–96. https://doi.org/10.1145/288392.288577

Peta Wyeth and Gordon Wyeth. 2001. Electronic Blocks: Tangible Programming Elements for Preschoolers. In Proceedings of the Eighth IFIP TC13 Conference on Human-Computer Interaction (INTERACT.

Ruhan Yang and Ellen Yi-Luen Do. 2024. PaBo Bot: Paper Box Robots for Everyone. In Companion of the 2024 ACM/IEEE International Conference on Human-Robot Interaction (HRI '24). Association for Computing Machinery, New York, NY, USA, 1158–1162. https://doi-org.colorado.idm.oclc.org/10.1145/3610978.3640696

Kentaro Yasu and Masahiko Inami. 2012. POPAPY: Instant Paper Craft Made Up in a Microwave Oven. In Advances in Computer Entertainment, Anton Nijholt, Teresa Romão and Dennis Reidsma (eds.). Springer Berlin Heidelberg, Berlin, Heidelberg, 406–420. https://doi.org/10.1007/978-3-642-34292-9_29

Iddo Yehoshua Wald and Oren Zuckerman. 2021. Magnetform: a Shape-change Display Toolkit for Material-oriented Designers. In Proceedings of the Fifteenth International Conference on Tangible, Embedded, and Embodied Interaction, 1–14. https://doi.org/10.1145/3430524.3446066

Clement Zheng, HyunJoo Oh, Laura Devendorf, and Ellen Yi-Luen Do. 2019. Sensing Kirigami. In Proceedings of the 2019 on Designing Interactive Systems Conference (DIS '19), 921–934. https://doi.org/10.1145/3322276.3323689

Clement Zheng, Peter Gyory, and Ellen Yi-Luen Do. 2020. Tangible Interfaces with Printed Paper Markers. In Proceedings of the 2020 ACM Designing Interactive Systems Conference. Association for Computing Machinery, New York, NY, USA, 909–923. https://doi.org/10.1145/3357236.3395578

Kening Zhu and Shengdong Zhao. 2013. AutoGami: a low-cost rapid prototyping toolkit for automated movable paper craft. In Proceedings of the SIGCHI Conference on Human Factors in Computing Systems (CHI '13), 661–670. https://doi.org/10.1145/2470654.2470748